\documentclass[amsmath,amssymb,12pt,superscriptaddress,nofootinbib]{revtex4-1}

\usepackage{graphicx}
\usepackage{dcolumn}
\usepackage{bm}
\usepackage{color}
\usepackage{enumitem}
  \usepackage{tabularx}
   \newcolumntype{C}{>{\centering\arraybackslash}X}
   \newcolumntype{L}{>{\raggedright\arraybackslash}X}
   \newcolumntype{R}{>{\raggedleft\arraybackslash}X}
\setlistdepth{10}

\newcommand{\dd}{\mathrm{d}}

\newcommand{\del}{\partial}
\newcommand{\ee}{{\rm e}}

\definecolor{DarkBlue}{rgb}{0,0,0.7} 

\definecolor{DarkRed}{rgb}{0.65,0,0} 
\newcommand{\dred}[1]{{#1}}

\begin{document}
\baselineskip5.5mm


{\baselineskip0pt
\small
\leftline{\baselineskip16pt\sl\vbox to0pt{
                             \vss}}
\rightline{\baselineskip16pt\rm\vbox to20pt{
\vspace{-1.5cm}
            \hbox{RUP-20-25}
            \hbox{KEK-Cosmo-261}
            \hbox{KEK-TH-2245}
\vss}}
}

\author{Chul-Moon~Yoo}\email{yoo@gravity.phys.nagoya-u.ac.jp}
\affiliation{
\fontsize{12pt}{1pt}\selectfont
Division of Particle and Astrophysical Science,
Graduate School of Science, \\Nagoya University, 
Nagoya 464-8602, Japan
\vspace{1.5mm}
}

\author{Tomohiro~Harada}\email{harada@rikkyo.ac.jp}
\affiliation{~
\fontsize{12pt}{1pt}\selectfont
Department of Physics, Rikkyo University, Toshima,
Tokyo 171-8501, Japan
\vspace{1.5mm}
}

\author{Shin'ichi~Hirano}\email{hirano.shinichi@a.mbox.nagoya-u.ac.jp}
\affiliation{
\fontsize{12pt}{1pt}\selectfont
Division of Particle and Astrophysical Science,
Graduate School of Science, \\Nagoya University, 
Nagoya 464-8602, Japan
\vspace{1.5mm}
}

\author{Kazunori~Kohri}\email{kohri@post.kek.jp}
\affiliation{
\fontsize{12pt}{1pt}\selectfont
Institute of Particle and Nuclear Studies, KEK, 1-1 Oho, Tsukuba, Ibaraki 305-0801, Japan
\vspace{1.5mm}
}
\affiliation{
\fontsize{12pt}{1pt}\selectfont
The Graduate University for Advanced Studies (SOKENDAI), 1-1 Oho, Tsukuba, Ibaraki 305-0801, Japan
\vspace{1mm}
}
\affiliation{
\fontsize{12pt}{1pt}\selectfont
Kavli Institute for the Physics and Mathematics of the
  Universe (WPI), University of Tokyo,
  Kashiwa 277-8583, Japan
\vspace{1mm}
}



\title{Abundance of Primordial Black Holes in Peak Theory \\for an Arbitrary Power Spectrum}


\begin{abstract}
\baselineskip5.5mm 
We modify the procedure to estimate PBH abundance proposed in Ref.~\cite{Yoo:2018kvb} 
so that 
it can be applied to a broad power spectrum such as the scale-invariant flat power spectrum. 
In the new procedure, we focus on peaks of the Laplacian of the curvature perturbation $\triangle \zeta$ 
and use the values of $\triangle \zeta$ and $\triangle \triangle \zeta $ at each peak to 
specify the profile of $\zeta$ as a function of the radial coordinate while 
the values of $\zeta$ and $\triangle \zeta$ are used in Ref.~\cite{Yoo:2018kvb}. 
The new procedure decouples the larger-scale environmental effect from the estimate of PBH abundance. 
Because the redundant variance due to the environmental effect is eliminated, 
we obtain a narrower shape of the mass spectrum 
compared to the previous procedure in Ref.~\cite{Yoo:2018kvb}. 
Furthermore, the new procedure 
allows us to estimate PBH abundance for the scale-invariant flat power spectrum 
by introducing a window function. 
Although the final result depends on the choice of the window function, 
we show that the $k$-space tophat window 
minimizes the extra reduction of the mass spectrum due to the window function. 
That is, the $k$-space tophat window has the minimum required property in the theoretical PBH estimation. 
Our procedure makes it possible to calculate the PBH mass spectrum 
for an arbitrary power spectrum by using a plausible PBH formation criterion 
with the nonlinear relation taken into account. 
\end{abstract}


\maketitle
\thispagestyle{empty}
\pagebreak

\section{Introduction}
\label{sec:intro}

Since Zel'dovich, Novikov and Hawking had pointed out its possibility\cite{1967SvA....10..602Z,Hawking:1971ei}, 
primordial black holes(PBHs) have continued to attract attention. 
They are still viable candidates for a substantial part of dark matter(see e.g. Refs.~\cite{Carr:2020gox,Carr:2020xqk} and references therein) and a possible origin of observed binary black holes\cite{Abbott:2016blz,Sasaki:2016jop}. 
Mass, spin or spatial distribution of PBHs provides us valuable information about relatively small scale inhomogeneity
in the early universe. 
When we connect a PBH production scenario and observational constraints on it, 
theoretical estimation of the PBH distribution is inevitable. 
Here, we provide a plausible procedure to calculate the PBH mass spectrum for an arbitrary power spectrum 
based on the peak theory. 

Until relatively recently,  
the Press-Schechter~(PS) formalism is applied to 
the estimation of PBH abundance based on a perturbation variable 
such as the comoving density or the curvature perturbation. 
As PBHs started to draw more attention after the discovery of the black hole binary as well as gravitational waves, 
people have begun 
to seriously doubt the relevance of the PS formalism in the estimation of PBH abundance. 
In order to improve the estimation, 
one needs to resolve the following mutually related issues: PBH formation criterion, 
statistical treatment of non-linear variables~\cite{DeLuca:2019qsy}, 
and use of a window function~\cite{Ando:2018qdb,Young:2019osy,Tokeshi:2020tjq}.  

For the criterion of the PBH formation, there has been a long-term debate since 
Carr had proposed a rough criterion\cite{Carr:1975qj}. 
A lot of efforts to clarify the appropriate criterion have been made through 
numerical and analytic treatments\cite{1978SvA....22..129N,1980SvA....24..147N,Shibata:1999zs,Niemeyer:1999ak,Musco:2004ak,Polnarev:2006aa,Musco:2008hv,Polnarev:2012bi,Nakama:2013ica,Harada:2013epa}. 
One useful criterion was proposed in Ref.~\cite{Shibata:1999zs} 
by using the compaction function, which is equivalent to the half of the 
volume average of the density perturbation in the long-wavelength limit\cite{Harada:2015yda}. 
In Ref.~\cite{Harada:2015yda}, through spherically symmetric numerical simulations, 
it was shown that the threshold of the maximum value of the 
compaction function gives relatively accurate criterion, which is within about 10\% accuracy for a moderate 
shape of initial configuration. 
More recently, the threshold value for the volume average of the compaction function was proposed in Ref.~\cite{Escriva:2019phb}, and 
it was shown that this variable gives the PBH formation criterion within 2\% accuracy 
for a moderate inhomogeneity in the radiation dominated universe(see Ref.~\cite{Escriva:2020tak} for 
general cosmological backgrounds). 
These recent developments show that the use of the compaction function is 
crucial for an accurate estimation of PBH abundance. 

Another important ingredient in the calculation of PBH abundance is 
the statistics of perturbation variables. 
Naively, we expect that the curvature perturbation would be relevant for the Gaussian 
distribution assumed in the PS formalism. 
However, the absolute value of the curvature perturbation 
does not have any physical meaning in a local sense because it can be 
absorbed into the coordinate rescaling.  
Therefore, setting the threshold value for the absolute value of 
the curvature perturbation seems irrelevant. 
Conversely, while setting the threshold on the compaction function would be appropriate, 
the compaction function cannot be a Gaussian variable even if the curvature perturbation is 
totally Gaussian because of their non-linear relation. 
Furthermore, difference of the gauge confuses the relation between perturbation variables. 

In Ref.~\cite{Harada:2015yda}, 
relations between different gauge conditions were summarized and 
the gauge issue has been clarified. 
The compaction function is expressed in terms of the curvature perturbation 
in the same reference~\cite{Harada:2015yda}. 
Then, apart from the window function, the remaining issue is how to count the number of PBHs 
by taking into account the nonlinear relation between the curvature perturbation 
and the compaction function. 
Although a few procedures treating the non-linear relation 
have been proposed\cite{Yoo:2018kvb,Germani:2019zez}, 
their consistency with each other has not been clear yet.

In Ref.~\cite{Yoo:2018kvb}, a plausible procedure to estimate PBH abundance for 
a narrow power spectrum of the Gaussian curvature perturbation was 
proposed, where the threshold for the compaction function is used and 
the non-linear relation is taken into account. 
However, this procedure cannot be directly applied to 
a broad spectrum%
(see Ref.~\cite{Suyama:2019npc} for a simple approach with linear relations). 
Our aim in this paper is to improve the procedure in Ref.~\cite{Yoo:2018kvb} so that 
we can introduce a window function, and make it possible to apply to any power spectrum.

This paper is organized as follows. 
First, the criterion based on the compaction function is introduced in Sec.~\ref{sec2}. 
In Sec.~\ref{peak_profile}, focusing on a high peak of the Laplacian of the curvature perturbation $\triangle \zeta$, 
we characterize the typical profile of the curvature perturbation $\zeta$ around the peak 
by using the values of $\triangle \zeta$ and $\triangle\triangle \zeta$. 
This treatment allows us to decouple the environmental effect on the absolute value of the curvature perturbation, 
and the criterion can be expressed in a purely local manner. 
In Sec.~\ref{PBH_abundance}, the procedure to estimate PBH abundance is explained, and 
applied to the single-scale narrow power spectrum previously presented in Ref.~\cite{Yoo:2018kvb}. 
We discuss the case of the scale-invariant flat spectrum implementing a window function in Sec.~\ref{window}. 
Sec.~\ref{discussion} is devoted to a summary and discussion.

Throughout this paper, we use the geometrized units in which both 
the speed of light and Newton's gravitational constant are unity, $c=G=1$.

\section{criterion for PBH formation}
\label{sec2}
Let us consider the 
spatial metric given by 
\begin{equation}
\dd s_3^2=a^2\ee^{-2\zeta}\tilde \gamma_{ij}\dd x^i \dd x^j 
\end{equation}
with $\det \tilde \gamma $ being the same as the determinant of the reference flat metric, 
where $a$ and $\zeta$ are the scale factor of the background universe and 
the curvature perturbation, respectively. 
In the long-wavelength approximation, 
the curvature perturbation $\zeta$ and the density perturbation $\delta$
with the comoving slicing are related by~\cite{Harada:2015yda},
\begin{equation}
\delta=-\frac{8}{9}\frac{1}{a^2H^2}\ee ^{5\zeta/2}\triangle\left(\ee^{-\zeta/2}\right) 
\end{equation}
in the radiation dominated universe, 
where $H$ is the Hubble expansion rate and 
$\triangle$ is the Laplacian of the reference flat metric. 

We will be interested mainly in high peaks, 
which tend to be nearly spherically symmetric~\cite{1986ApJ...304...15B}.
Therefore, in this section, we introduce the criterion for PBH formation originally proposed in Ref.~\cite{Shibata:1999zs} assuming spherical symmetry. 
Here, we basically follow and refer to the discussions and calculation in Ref.~\cite{Harada:2015yda}. 

First, let us define the compaction function $\mathcal C$ as 
\begin{equation}
\mathcal C:=\frac{\delta M}{R}, 
\end{equation}
where $R$ is the areal radius at the radius $r$, 
and $\delta M$ is the excess of the Misner-Sharp mass 
enclosed by the sphere of the radius $r$ compared with the 
mass inside the sphere in the fiducial flat Friedmann-Lema\^itre-Robertson-Walker universe 
with the same areal radius. 
From the definition of $\mathcal C$, we can derive the following simple form 
in the comoving slicing~(see also Eq.~(6.33) in Ref.~\cite{Harada:2015yda}):
\begin{equation}
\mathcal C(r)=\frac{1}{3}\left[1-\left(1-r\zeta'\right)^2\right]. 
\label{eq:Compaction}
\end{equation}

We will 
assume that the function $\mathcal C$ is a smooth function of $r$ for $r>0$. 
Then, the value of $\mathcal C$ takes the maximum value $\mathcal C^{\rm max}$ at $r_{\rm m}$ 
which satisfies
\begin{equation}
\mathcal C'(r_{\rm m})=0\Leftrightarrow \left(\zeta'+r\zeta''\right)|_{r=r_{\rm m}}=0. 
\label{eq:forrm}
\end{equation}
We consider the following criterion for PBH formation:
\begin{equation}
\mathcal C^{\rm max}>\mathcal C_{\rm th} \equiv \frac{1}{2}\delta_{\rm th}. \label{maxth}
\end{equation}
In the comoving slicing, the threshold $\mathcal C_{\rm th}$ for PBH formation 
is evaluated as $\simeq0.267$ (see Figs.~2 and 3 or TABLE I and II in Ref.~\cite{Harada:2015yda}). 
This threshold corresponds to the perturbation profiles of Refs.~\cite{Shibata:1999zs,Polnarev:2006aa,Musco:2012au}, 
and is found to be quite robust for a broad range of parameters(see Ref.~\cite{Escriva:2019phb} for a 
more robust criterion). 
In this paper we shall use this value as a reference value.

\section{Peak of $\triangle \zeta$ and the spherical profile}
\label{peak_profile}

Throughout this paper, we assume the random Gaussian distribution of $\zeta$ with
its power spectrum $\mathcal P(k)$ defined by the following equation:
\begin{equation}
<\tilde \zeta^*(\bm k)\tilde \zeta(\bm k')>=\frac{2\pi^2}{k^3}\mathcal P(k)(2\pi)^3\delta(\bm k-\bm k'), 
\label{eq:zeta_power}
\end{equation}
where $\tilde \zeta(\bm k)$ is the Fourier transform of $\zeta$ and 
the bracket $<...>$ denotes the ensemble average. 
Each gradient moment $\sigma_n$ can be calculated by 
\begin{equation}
\sigma_n^2:=\int \frac{\dd k}{k}k^{2n} \mathcal  P(k). 
\end{equation}
Hereafter we suppose that the power spectrum is given. 
Then the gradient moments can be calculated from the 
power spectrum and regarded as constants.

In this paper, we focus on high peaks of $\zeta_2:=\triangle \zeta$, which coincide with 
peaks of $\delta$ with linear relation. 
We note that this procedure is different from the previous one proposed in Ref.~\cite{Yoo:2018kvb}, 
where peaks of $-\zeta$ is considered. 

Focusing on a high peak of $\zeta_2$ 
and taking it as the origin of the coordinates, 
we introduce the amplitude $\mu_2$ and the curvature scale $1/k_\bullet$ of the peak as follows:%
\footnote{
\baselineskip5mm
Notations of the amplitude $\mu_2$ and the curvature scale $k_\bullet$ are chosen 
so that they will be distinguished from $\mu$ and $k_*$ in Ref.~\cite{Yoo:2018kvb}. }
\begin{eqnarray}
\mu_2&=&\left. \zeta_2\right|_{r=0}, 
\label{eq:mudef}\\
k_\bullet^2&=&-\frac{\left.\triangle \triangle \zeta\right|_{r=0}}{\mu_2}. 
\label{eq:ksdef}
\end{eqnarray}
According to the peak theory\cite{1986ApJ...304...15B}, 
for a high peak, assuming the spherical symmetry, 
we may expect the typical form of the profile $\bar \zeta_2$ can be described by using 
$\mu_2$ and  $k_\bullet$ 
as follows:
\begin{equation}
\frac{\bar \zeta_2(r)}{\sigma_2}=\frac{\mu_2/\sigma_2}{1-\gamma_3^2}\left(\psi_2+\frac{1}{3}R_\bullet^2 \triangle\psi_2\right)-\frac{\mu_2k_\bullet^2/\sigma_4}{\gamma_3(1-\gamma_3^2)}\left(\gamma_3^2\psi_2+\frac{1}{3}R_\bullet^2 \triangle\psi_2\right), 
\label{eq:zetahat}
\end{equation}
with $\gamma_3=\sigma_3^2/(\sigma_2\sigma_4)$, $R_\bullet=\sqrt{3}\sigma_3/\sigma_4$ and 
\begin{equation}
\psi_n(r)=\frac{1}{\sigma_2^2}\int \frac{\dd k}{k} k^{2n} \frac{\sin(kr)}{kr}\mathcal P(k). 
\label{eq:psi}
\end{equation}
It is worthy of note that, for $k_\bullet=\sigma_3/\sigma_2$, 
we obtain 
\begin{equation}
\bar \zeta_2(r;\sigma_3/\sigma_2)=\mu_2\psi_2(r). 
\label{eq:gkc}
\end{equation}
It will be shown in Eq.~\eqref{eq:P1} that 
regarding $k_\bullet$ as a probability variable, we obtain $\sigma_3/\sigma_2$ as the mean value of $k_\bullet$. 

Let us consider the profile $\bar \zeta$ given by integrating Eq.~\eqref{eq:zetahat}. 
Integrating $\bar \zeta_2$, and assuming the regularity at the center, we obtain 
\begin{equation}
\frac{\bar \zeta(r)}{\sigma_2}=-\frac{\mu_2/\sigma_2}{(1-\gamma_3^2)}\left(\psi_1+\frac{1}{3}R_\bullet \triangle\psi_1\right)+\frac{\mu_2k_\bullet^2/\sigma_4}{\gamma_3(1-\gamma_3^2)}\left(\gamma_3^2\psi_1+\frac{1}{3}R_\bullet \triangle\psi_1\right)+\frac{\zeta_\infty}{\sigma_2}, 
\end{equation}
where $\zeta_\infty=\bar \zeta|_{r=\infty}$ is an integration constant. 
Because we have $\psi_1|_{r=0}=\sigma_1^2/\sigma_2^2$ and $\triangle \psi_1|_{r=0}=-1$, 
we obtain 
\begin{equation}
\zeta_0:=\bar \zeta|_{r=0}=-\mu_2\frac{\sigma_1^2\sigma_4^2-\sigma_2^2\sigma_3^2+(\sigma_2^4-\sigma_1^2\sigma_3^2)k_\bullet^2}{\sigma_2^2\sigma_4^2-\sigma_3^4}+\zeta_\infty. 
\label{eq:zeta0inf}
\end{equation}
We may consider either $\zeta_0$ or $\zeta_\infty$ as a probability variable. 
Since the constant shift of $\zeta$ can be absorbed into the 
renormalization of the background scale factor, 
the non-zero value of $\zeta_\infty$ would be regarded as 
a larger-scale environmental effect. 
Actually, in Appendix\ref{Gauss}, we show that the mean value of $\zeta_\infty$ is 0 
for a given set of $\mu_2$ and $k_\bullet$. 
In Ref.~\cite{Yoo:2018kvb}, we used $\zeta_0$ to characterize the profile of the 
curvature perturbation. 
However, the use of $\zeta_0$ would mix the environmental effect with the local state. 
Therefore, in this paper, we ignore $\zeta_\infty$ by renormalizing the background scale factor as 
$\ee^{-\zeta_\infty}a\rightarrow a$ to eliminate the environmental effect, and 
regard $\zeta_0$ as a dependent variable on $\mu_2$ and $k_\bullet$ through Eq.~\eqref{eq:zeta0inf} 
with $\zeta_\infty=0$.

In order to obtain PBH abundance, we can follow the procedure proposed in Ref.~\cite{Yoo:2018kvb}
replacing $\mu$ and $k_*$ by $\mu_2$ and $k_\bullet$. 
Here, we just copy the part of the procedure from Ref.~\cite{Yoo:2018kvb}(the flow char of our procedure can be 
seen in Ref.~\cite{Yoo:2019pma}). 
Applying Eq.~\eqref{eq:Compaction} to $\bar\zeta$, we obtain the relation between $\mu_2$ and $\mathcal C$ as 
\begin{equation}
\mu_2=\frac{1-\sqrt{1-3\mathcal C}}{rg'}, 
\end{equation}
where $g(r;k_\bullet):=\bar\zeta/\mu_2$ and the smaller root is taken. 
Let us define the threshold value $\mu_{\rm 2th}^{(k_\bullet)}$ as 
\begin{equation}
\mu_{\rm 2th}^{(k_\bullet)}(k_\bullet )=\frac{1-\sqrt{1-3\mathcal C_{\rm th}}}{\bar r_{\rm m}(k_\bullet)g_{\rm m}'(k_\bullet)}, 
\label{eq:muth}
\end{equation}
where 
$\bar r_{\rm m}(k_\bullet)$ is the value of $r_{\rm m}$ for $\zeta=\bar \zeta$, and
\begin{equation}
g_{\rm m}(k_\bullet):=g(\bar r_{\rm m}(k_\bullet);k_\bullet). 
\end{equation}
In Eq.~\eqref{eq:muth}, we explicitly denoted the $k_\bullet$ dependence of 
$\bar r_{\rm m}$ and $g_{\rm m}$ to emphasize it. 

In order to express the threshold value as a function of the PBH mass $M$, 
let us consider the horizon entry condition: 
\begin{equation}
aH=\frac{a}{R(\bar r_{\rm m})}=\frac{1}{\bar r_{\rm m}}\ee^{\mu_2 g_{\rm m}}.  \label{hentry}
\end{equation}
Since the PBH mass is
given by $M=\alpha/(2H)$ with $\alpha$ being a numerical factor, from
the horizon entry condition (\ref{hentry}), the PBH mass $M$ can be expressed as
follows:
\begin{eqnarray}
M&=&\frac{1}{2}\alpha H^{-1}=\frac{1}{2}\alpha a\bar r_{\rm m}\ee^{-\mu_2 g_{\rm m}}
=M_{\rm eq}k_{\rm eq}^2\bar r_{\rm m}^2\ee^{-2\mu_2 g_{\rm m}}=:M^{(\mu_2,k_\bullet)}(\mu_2,k_\bullet), 
\label{eq:kM}
\end{eqnarray}
where we have used the fact $H\propto a^{-2}$ and
$a =a_{\rm eq}^2 H_{\rm eq} \bar r_{\rm m}\ee^{-\mu_2 g_{\rm m}}$ 
with $a_{\rm eq}$ and
$H_{\rm eq}$ being the scale factor and Hubble expansion rate at the
matter-radiation equality.  
$M_{\rm eq}$ and $k_{\rm eq}$ are defined
by $M_{\rm eq}=\alpha H_{\rm eq}^{-1}/2$ and
$k_{\rm eq}=a_{\rm eq}H_{\rm eq}$, respectively. 
For simplicity, we 
set $\alpha=1$ as a fiducial value%
\footnote{
\baselineskip5mm
In order to take into account the critical 
behavior\cite{Choptuik:1992jv,Koike:1995jm}, 
$\alpha$ should be given by a function of $\mu_2$ and $k_\bullet$ as 
$\alpha= K(k_\bullet)(\mu_2-\mu_{\rm 2th}(k_\bullet))^\gamma$ with $\gamma\simeq 0.36$~\cite{Niemeyer:1997mt,Yokoyama:1998xd,Niemeyer:1999ak,Green:1999xm,Musco:2004ak,Musco:2008hv,Musco:2012au,Kuhnel:2015vtw,Germani:2018jgr} 
and $K(k_\bullet)$ being some function of $k_\bullet$, which would be profile dependent.  
}. 

Then we can obtain the threshold value of $\mu_{\rm 2th}^{(M)}(M)$ 
as a function of $M$ by eliminating $k_\bullet$ from 
Eqs.~\eqref{eq:kM} and $\mu_2=\mu^{(k_*)}_{\rm 2th}(k_\bullet)$, and 
solving it for $\mu_2$. 
That is, defining 
$k_\bullet^{\rm th}(M)$ by the inverse function of 
$M=M^{(\mu_2,k_\bullet)}(\mu_{\rm 2th}^{(k_\bullet)}(k_\bullet),k_\bullet)$, 
we obtain the threshold value of $\mu_{\rm 2th}^{(M)}$ for a fixed value of $M$ as 
\begin{equation}
\mu_{\rm 2th}^{(M)}(M):=\mu_{\rm 2th}^{(k_\bullet)}(k_\bullet^{\rm th}(M)). 
\label{eq:muthM}
\end{equation}
While, from Eq.~\eqref{eq:kM}, we can describe $\mu_2$ as a function of $M$ and $k_\bullet$ as follows:
\begin{equation}
\mu_2=
\mu^{(M,k_\bullet)}(M,k_\bullet):=
-\frac{1}{2g_{\rm m}}\ln \left(\frac{1}{k_{\rm eq}^2\bar r_{\rm m}^2}\frac{M}{M_{\rm eq}}\right). 
\end{equation}
The value of $\mu_2$ may be bounded below 
by $\mu_{\rm 2min}(M)$ for a fixed value of $M$. 
Actually, in the specific examples in Sec.~\ref{PBH_abundance} and \ref{window}, 
the value of $\mu_{\rm 2min}(M)$ is given as follows:
\begin{equation}
\mu_{\rm 2min}=\mu^{(M,k_\bullet)}(M,0). 
\label{eq:mu2min}
\end{equation}
Then, for a fixed value of $M$, the region of $\mu$ for PBH formation can be given by 
\begin{equation}
\mu_2>\mu_{\rm 2b}:=\max\left\{\mu_{\rm 2min}(M),\mu_{\rm 2th}^{(M)}(M)\right\}. 
\label{eq:mub}
\end{equation}

\section{PBH abundance}
\label{PBH_abundance}
From Ref.~\cite{Yoo:2018kvb}, 
we obtain the expression for the peak number density characterized by $\mu_2$ and $k_\bullet$
as 
\begin{eqnarray}
&&n^{(k_\bullet)}_{\rm pk}(\mu_2, k_\bullet) \dd \mu_2\dd k_\bullet
=\dred{\frac{2 }{3^{3/2}(2\pi)^{3/2}}}\mu_2 k_\bullet\frac{\sigma_4^2}{\sigma_2\sigma_3^3}f\left(\frac{\mu_2 k_\bullet^2}{\sigma_4}\right)
P_1\left(\frac{\mu_2}{\sigma_2},\frac{\mu_2 k_\bullet^2}{\sigma_4}\right) 
\dd\mu_2\dd k_\bullet,   
\label{eq:nks}
\end{eqnarray}
where 
\begin{eqnarray}
f(x)&=&
\frac{1}{2}x(x^2-3)
\left({\rm erf}\left[\frac{1}{2}\sqrt{\frac{5}{2}}x\right]
+{\rm erf}\left[\sqrt{\frac{5}{2}}x\right]\right)\cr
&&\hspace{1cm}+\sqrt{\frac{2}{5\pi}}\left\{
\left(\frac{8}{5}+\frac{31}{4}x^2\right)
\exp\left[-\frac{5}{8}x^2\right]
+\left(-\frac{8}{5}+\frac{1}{2}x^2\right)\exp\left[-\frac{5}{2}x^2\right]
\right\}, 
\label{eq:funcf}
\end{eqnarray}
and 
\dred{
  \begin{equation}
P_1 \left( \frac{\mu_2}{\sigma_2},\frac{\mu_2 k_\bullet^2}{\sigma_4} \right) 
=\frac{1}{2\pi\sqrt{1-\gamma_3^2}}
\exp \left[
-\frac{\mu_2^2}{2{\tilde \sigma(k_\bullet)}^2} 
\right]
\label{eq:P1}
\end{equation}
}
with 
\begin{equation}
\frac{1}{\tilde \sigma^2(k_\bullet)}:=\frac{1}{\sigma_2^2}+\frac{1}{\sigma_4^2(1-\gamma_3^2)}\left(k_\bullet^2-\frac{\sigma_3^2}{\sigma_2^2}\right)^2. 
\label{eq:deltil}
\end{equation}
In Eq.~\eqref{eq:nks}, 
the following replacements have been made from Eq.~(58) in Ref.~\cite{Yoo:2018kvb}:
\begin{equation}
\mu\rightarrow \mu_2,~k_*\rightarrow k_\bullet,~\sigma_n\rightarrow\sigma_{n+2},~\gamma\rightarrow \gamma_3. 
\end{equation}
Since the direct observable is not $k_\bullet$ but the PBH mass $M$, we
further change the variable from $k_\bullet$ to $M$ as follows:
\begin{eqnarray}
n^{(M)}_{\rm pk}(\mu_2, M)\dd \mu_2\dd M
&:=&n^{(k_\bullet)}_{\rm pk}(\mu_2, k_\bullet)\dd \mu_2\dd k_\bullet\cr
&=&\dred{3^{-3/2}(2\pi)^{-3/2}}
\frac{\sigma_4^2}{\sigma_2\sigma_3^3}
\mu_2 k_\bullet
f\left(\frac{\mu_2 k_\bullet^2}{\sigma_4}\right)\cr
&&\hspace{1cm}P_1\left(\frac{\mu_2}{\sigma_2},\frac{\mu_2 k_\bullet^2}{\sigma_4}\right) 
\left|\frac{\dd}{\dd k_\bullet}\ln \bar r_{\rm m}-\mu_2 \frac{\dd}{\dd k_\bullet}g_{\rm m}\right|^{-1}
\dd\mu_2\dd \ln M,  
\end{eqnarray}
where $k_\bullet$ should be regarded as a function of $\mu_2$ and $M$ 
given by solving Eq.~\eqref{eq:kM} for $k_\bullet$. 
We note that an extended power spectrum is implicitly assumed in the above expression. 
In the monochromatic spectrum case, the expression reduces to the same expression in Ref.~\cite{Yoo:2018kvb}
because $\sigma_n=\sigma$ for $\mathcal P(k)=\sigma^2 k_0\delta(k-k_0)$. 

It should be noted that, since we relate $k_\bullet$ to $M$ with $\mu_2$ fixed, 
we have implicitly assumed that there is only one peak with $\triangle \zeta_2=-\mu_2 k_\bullet^2$ 
in the region corresponding to the mass $M$, that is, inside $r=r_{\rm m}$. 
If the spectrum is broad enough or has multiple peaks at far-separated scales, 
and the typical PBH mass is relatively larger than the minimum scale given by the spectrum, 
we would find multiple peaks inside $r=r_{\rm m}$. 
Then we cannot correctly count the number of peaks 
in the scale of interest. 
In order to avoid this difficulty, 
we need to introduce a window function to smooth out the smaller-scale inhomogeneities. 
We discuss this issue in the subsequent section. 
In this section, we simply assume that the power spectrum is characterized by a single scale $k_0$ and 
there is no contribution from the much smaller scales $k\gg k_0$.

The number density of PBHs is given by 
\begin{equation}
n_{\rm BH} \dd \ln M=\left[
\int^\infty_{\mu_{\rm 2b}}\dd \mu_2
~n^{(M)}_{\rm pk}(\mu_2,M)\right]M \dd \ln M. 
\end{equation}
We also note that the scale factor $a$ is a function of $M$ as $a=2M^{1/2}M_{\rm eq}^{1/2}k_{\rm eq}/\alpha$. 
Then the fraction of PBHs to the total density $\beta_0\dd \ln M$ can be given by 
\begin{eqnarray}
\beta_0\dd \ln M&=&\frac{M n_{\rm BH}}{\rho a^3}\dd \ln M
=\frac{4\pi}{3}\alpha n_{\rm BH}k_{\rm eq}^{-3}\left(\frac{M}{M_{\rm eq}}\right)^{3/2} \dd \ln M\cr
&=&\dred{\frac{2\alpha k_{\rm eq}^{-3}}{3^{5/2}(2\pi)^{1/2}}}
\frac{\sigma_4^2}{\sigma_2\sigma_3^3}
\left(\frac{M}{M_{\rm eq}}\right)^{3/2}
\Biggl[
\int^\infty_{\mu_{\rm 2b}}\dd\mu 
\mu_2 k_\bullet f\left(\frac{\mu_2 k_\bullet^2}{\sigma_4}\right)
\cr
&&\hspace{2cm}P_1\left(\frac{\mu_2}{\sigma_2},\frac{\mu_2 k_\bullet^2}{\sigma_4}\right)
\left|\frac{\dd}{\dd k_\bullet}\ln \bar r_{\rm m}-\mu_2 \frac{\dd}{\dd k_\bullet}g_{\rm m}\right|^{-1}
\Biggr]\dd \ln M. 
\label{eq:beta_general}
\end{eqnarray}
Here we note again that $k_\bullet$ should be regarded as a function of $\mu_2$ and $M$. 
The above formula can be evaluated in principle once the form of the power spectrum is given. 
A crucial difference of Eq.~\eqref{eq:beta_general} from Eq.~(61) in Ref.~\cite{Yoo:2018kvb} is 
that the expression does not depend on $\sigma_0$, which has IR-log divergence for the flat scale-invariant spectrum. 
Thus we can consider the PBH mass spectrum for the flat spectrum without introducing 
IR cut-off. 
\footnote{
\baselineskip5mm
The PBH fraction to the total density $f_0$ at the equality time is given by $f_0=\beta_0 (M_{\rm eq}/M)^{1/2}$. 
We do not explicitly show the form of $f_0$ in this paper since the scale dependence can be more 
easily understood by the form of $\beta_0$. 
}

In order to give a simpler approximate form of Eq.~\eqref{eq:beta_general}, 
we approximately perform the integral with respect to $\mu$ as follows:
\begin{eqnarray}
\beta_0\dd \ln M&\simeq&
\dred{\frac{2\alpha k_{\rm eq}^{-3}}{3^{5/2}(2\pi)^{1/2}}}
\frac{\sigma_4^2}{\sigma_2\sigma_3^3}
\left(\frac{M}{M_{\rm eq}}\right)^{3/2}
\Biggl[
\tilde \sigma(k_\bullet)^2
k_\bullet f\left(\frac{\mu_2 k_\bullet^2}{\sigma_4}\right)
\cr
&&\hspace{2cm}P_1\left(\frac{\mu_2}{\sigma_2},\frac{\mu_2 k_\bullet^2}{\sigma_4}\right)
\left|\frac{\dd}{\dd k_\bullet}\ln \bar r_{\rm m}-\mu_2 \frac{\dd}{\dd k_\bullet}g_{\rm m}\right|^{-1}
\Biggr]_{\mu_2=\mu_{\rm 2b}}\dd \ln M. 
\label{eq:beta_approx}
\end{eqnarray}
Since $P_1$ given in Eq.~\eqref{eq:P1} has the exponential dependence, 
we may expect that the value of $\beta_0$ is sensitive to the exponent $-\mu_2^2/2\tilde \sigma^2$.
Therefore, 
assuming $\mu_{\rm 2b}=\mu_{\rm 2th}^{(M)}=\mu_{\rm 2th}^{(k_\bullet)}(k_\bullet^{\rm th}(M))$, we can roughly estimate the maximum value of $\beta_0$ at the top of the mass spectrum 
by considering the value $k_{\rm t}$ of $k_\bullet$ which minimizes the value of $\mu_{\rm 2th}^{k_\bullet}(k_\bullet)/\tilde \sigma$,%
\footnote{
\baselineskip5mm
The expansion around $k_\bullet=\sigma_3/\sigma_2$ 
like in Ref.~\cite{Yoo:2018kvb} 
does not work well because of the large $k_\bullet$ dependence of $\mu_{\rm 2th}^{(k_\bullet)}$. 
That is, the peak of the exponent $-\mu_{\rm 2th}^2/\tilde \sigma^2$ 
significantly deviates from $k_\bullet=\sigma_3/\sigma_2$ and the Taylor expansion is not as effective as 
in Ref.~\cite{Yoo:2018kvb}.  
} namely, 
\begin{equation}
k_{\rm t}:={\rm argmin}_{k_\bullet} \left[\mu_{\rm 2th}^{(k_\bullet)}(k_\bullet)/\tilde \sigma(k_\bullet)\right].  
\footnote{
\baselineskip5mm
${\rm argmin}_x f(x)=\{x~|~\forall y (f(y) 
\geq f(x))\}$. 
}
\end{equation}
The value of $k_{\rm t}$ cannot be given in an analytic form in general, and 
a numerical procedure to find the value of $k_{\rm t}$ is needed. 
We note that the value of $k_{\rm t}$ is independent of the overall constant factor of the power spectrum, and 
depends only on the profile of the spectrum. 
Substituting $k_{\rm t}$ into $k_\bullet$ in Eq.~\eqref{eq:beta_approx}, 
we obtain the following rough estimate for the maximum value $\beta_{\rm 0,max}$: 
\begin{eqnarray}
\beta_{\rm 0,max}&\simeq&
\beta_{\rm 0,max}^{\rm approx}
:=
\dred{\frac{2\alpha k_{\rm eq}^{-3}}{3^{5/2}(2\pi)^{1/2}}}
\frac{\sigma_4^2}{\sigma_2\sigma_3^3}
\left(\frac{M_{\rm t}}{M_{\rm eq}}\right)^{3/2}
\Biggl[
\tilde \sigma(k_\bullet)^2
k_\bullet f\left(\frac{\mu_2 k_\bullet^2}{\sigma_4}\right)
\cr
&&\hspace{2cm}P_1\left(\frac{\mu_2}{\sigma_2},\frac{\mu_2 k_\bullet^2}{\sigma_4}\right)
\left|\frac{\dd}{\dd k_\bullet}\ln \bar r_{\rm m}-\mu_2 \frac{\dd}{\dd k_\bullet}g_{\rm m}\right|^{-1}
\Biggr]_{k_\bullet=k_{\rm t},~\mu_2=\mu_{\rm 2th}^{(k_\bullet)}(k_{\rm t})}, 
\label{eq:beta_max}
\end{eqnarray}
where $M_{\rm t}:=M^{(\mu_2,k_\bullet)}(\mu_{\rm 2th}^{(k_\bullet)}(k_{\rm t}), k_{\rm t})$. 
We note that 
the expression~\eqref{eq:beta_max} 
gives a better approximation for a smaller value of the amplitude of the power spectrum $\sigma^2$, and 
may have a factor of difference from the actual maximum value for $\sigma\gtrsim0.1$ 
due to the mass dependence of the factors other than the exponential part 
as can be found in the examples below. 

\begin{figure}[htbp]
\begin{center}
\includegraphics[scale=0.65]{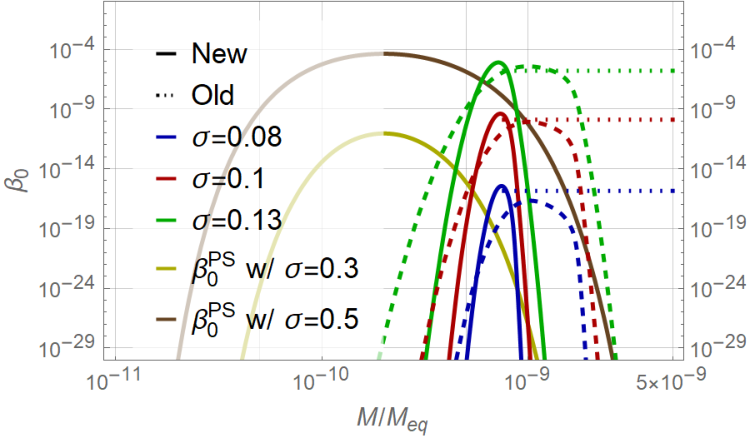}
\includegraphics[scale=0.62]{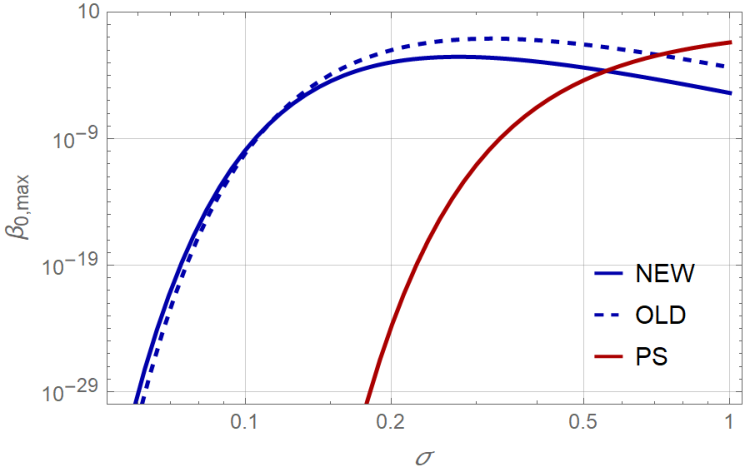}
\caption{
\baselineskip5mm
PBH mass spectrum(left) and $\beta^{\rm approx}_{\rm 0,max}$
as a function of $\sigma$(right) 
for the extended power spectrum \eqref{eq:Gausspower} with $k_0=10^5k_{\rm eq}$. 
The solid lines correspond to the spectra calculated by our new procedure with $\alpha=1$, and 
the dashed lines show the spectra calculated in Ref.~\cite{Yoo:2018kvb}. 
We also plot the mass spectrum $\beta_0^{\rm PS}$ obtained from the Press-Schechter formalism explained in Appendix 
for comparison. 
In the left panel, the dotted horizontal lines show the corresponding values of 
$\beta_{\rm 0,max}^{\rm approx}$.  
}
\label{fig:single_scale}
\end{center}
\end{figure}

Let us consider the extended power spectrum given by 
\begin{equation}
\mathcal  P(k)=3\sqrt{\frac{6}{\pi}}\sigma^2\left(\frac{k}{k_0}\right)^3
\exp\left(-\frac{3}{2}\frac{k^2}{k_0^2}\right). 
\label{eq:Gausspower}
\end{equation}
Gradient moments are calculated as 
\begin{equation}
\sigma_n^2=\frac{2^{n+1}}{3^n\sqrt{\pi}}\Gamma\left(\frac{3}{2}+n\right)\sigma^2k_0^{2n}, 
\end{equation}
where $\Gamma$ means the gamma function. 
The result is shown in Fig.~\ref{fig:single_scale}. 
Our new procedure gives a narrower and slightly higher spectrum than that 
obtained in Ref~\cite{Yoo:2018kvb}.
This behavior can be understood as the environmental effect induced by 
the variance of $\zeta_\infty$. 
Although the effect is so small that it could be practically ignored in this example, 
we successfully decoupled the environmental effect. 

\section{Implementing a window function}
\label{window}
In our new procedure, 
a window function can be straightforwardly implemented. 
That is, introducing the UV cut-off scale $k_W$, instead of Eq.~\eqref{eq:zeta_power}, we consider 
the following power spectrum of $\zeta$:
\begin{equation}
\mathcal P_W(k)=\mathcal P(k)W(k;k_W)^2, 
\end{equation}
where $W(k;k_W)$ is a window function satisfying $W(k;k_W)\leq1$ and $W(k;k_W)=0$ for $k\gg k_W$. 
Then, following the procedure given in the previous section, we can calculate 
PBH abundance for a given value of $k_W$. 
The final PBH mass spectrum is given by the envelope curve of the mass spectra for all values of $k_W$. 
We note that, for a narrow power spectrum, $\mathcal P_W(k)\rightarrow\mathcal P(k)$ 
in the limit $k_W\rightarrow\infty$, and 
the mass spectrum results in the case without the window function irrespective of the choice of the window function.

\begin{figure}[htbp]
\begin{center}
\includegraphics[scale=0.8]{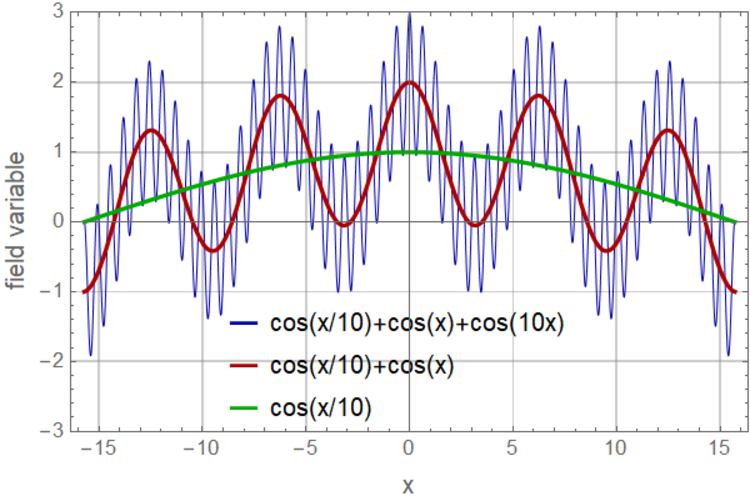}
\caption{
\baselineskip5mm
Three functions, $\cos(x/10)+\cos(x)+\cos(10x)$, $\cos(x/10)+\cos(x)$ and $\cos(x/10)$ 
are plotted as functions of $x$. 
For $\cos(x/10)+\cos(x)$, every peak has the scale of order 1, but 
the peak profiles are sharper for $\cos(x/10)+\cos(x)+\cos(10x)$ and broader for $\cos(x/10)$. 
}
\label{fig:coss}
\end{center}
\end{figure}

One important issue here is the window function dependence of the final mass spectrum. 
In order to clarify this issue, for a sufficiently broad power spectrum, let us consider 
the effect of the window function for a peak number density at a fixed scale given by the wave number $k_\bullet=k_0$. 
If $k_0\gg k_W$, no peak can be found. 
On the other hand, if $k_0\ll k_W$, we would find many smaller scale peaks inside 
the region of the radius $\sim 1/k_0$ because the smaller scale modes with $k\gg k_0$ are 
superposed on top of the inhomogeneity with $k\sim k_0$. 
Thus 
every peak 
has a sharp profile due to the superposed small-scale inhomogeneity 
and satisfies $k_\bullet\gg k_0$. 
This means that there is essentially no peak with $k_\bullet=k_0\ll k_W$ 
if $k_0\ll k_W$ and the original power spectrum has a 
sufficiently broad support in $k> k_0$(Fig.~\ref{fig:coss} would be helpful for understanding). 
For a fixed $k_\bullet=k_0$, in the both limits $k_0\ll k_W$ and $k_0\gg k_W$, 
the number of peaks decreases. 

In our procedure, 
since we take the envelope curve for all values of $k_W$, 
the final estimate for the peak number density at $k_\bullet = k_0$ is given by the value 
for $k_W$ which maximizes the peak number density at $k_\bullet =k_0$. 
For this specific value of $k_W$, $k_0$ corresponds to $k_{\rm t}$ introduced in 
Eq.~\eqref{eq:beta_max}, which basically maximizes the peak number, 
and generally we have $k_W>k_{\rm t}\simeq k_0$. 
If the window function reduces the amplitude of the power spectrum in the region of $k$ 
much smaller than $k_W$, the maximum number density of peaks with $k_\bullet = k_0\simeq k_{\rm t}<k_W$ 
also inevitably decreases due to the window function. 
Thus the final estimate for the peak number density at $k_\bullet = k_0$ also decreases. 
For this reason,  we expect that a sharp cut-off of the window function would 
provide a larger value of the peak number density minimizing the extra reduction of the mass spectrum 
due to the window function.

\begin{figure}[htbp]
\begin{center}
\includegraphics[scale=0.65]{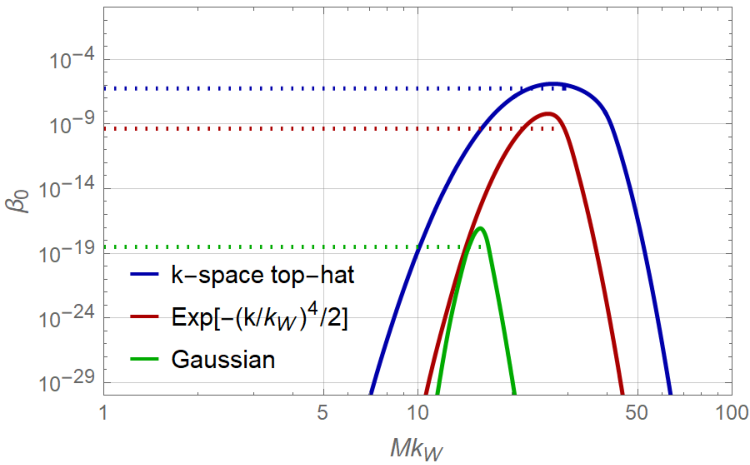}
\includegraphics[scale=0.62]{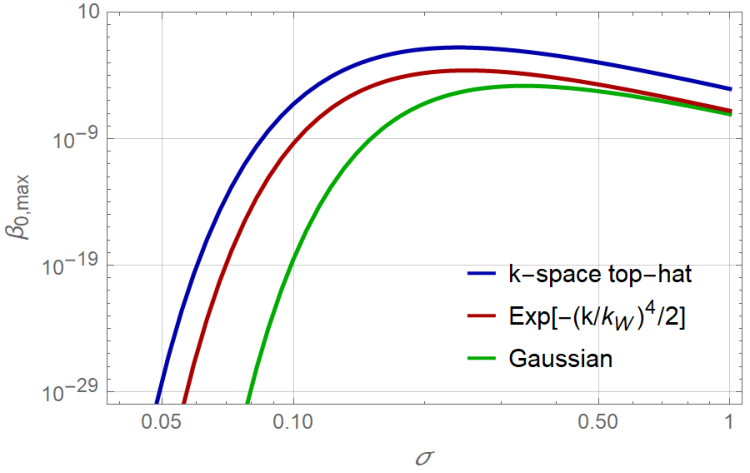}
\caption{
\baselineskip5mm
PBH mass spectrum(left) and $\beta_{\rm 0,max}^{\rm approx}$ 
as a function of $\sigma$(right) 
for the flat power spectrum with each window function. 
In the left panel, we set $\sigma=0.1$ and 
the dotted horizontal lines show the corresponding values of $\beta_{\rm 0,max}^{\rm approx}$. 
}
\label{fig:winddep}
\end{center}
\end{figure}

Let us check the above discussion by considering the flat scale-invariant spectrum with a window function:
\begin{equation}
\mathcal P_W(k;k_W)=\sigma^2 W(k;k_W)^2.  
\end{equation}
We consider the following window functions:
\begin{eqnarray}
W_n(k/k_W)&=&\exp\left(-\frac{1}{2}\frac{k^{2n}}{k_W^{2n}}\right),\\ 
W_{k{\rm TH}}(k/k_W)&=&\Theta(k_W-k), 
\label{eq:windows}
\end{eqnarray}
where we note that $W_1$ is the standard Gaussian window function%
\footnote{
\baselineskip5mm
The real-space tophat window function leads divergent gradient moments 
for the scale-invariant flat spectrum, so that 
we practically cannot use it. 
}. 
For each window function, we can calculate the PBH mass spectrum with a fixed value of $k_W$ 
following the procedure presented in the previous section. 
The results are shown in Fig.~\ref{fig:winddep}. 
As is shown in Fig.~\ref{fig:winddep}, the result significantly depends on the window function. 
For the overall mass spectrum, taking the envelope curve of the mass spectra for all values of $k_W$, 
we obtain the flat mass spectrum with the amplitude given by the maximum value in 
the plot. 
Therefore the $k$-space tophat window function gives the largest abundance as is expected. 
This behavior is contrary to the case of the conventional Press-Schechter~(PS) formalism 
where the Gaussian window function gives a larger abundance than that for the $k$-space tophat window(see Appendix~\ref{sec:PS} and Ref.~\cite{Ando:2018qdb}). 
The $\sigma$-dependence of $\beta_{\rm 0,max}^{\rm approx}$, which gives an order-of-magnitude estimate for 
the maximum value of the mass spectrum, 
is also shown in Fig.~\ref{fig:winddep}.

\section{Summary and discussion}
\label{discussion}

In this paper, we have improved the procedure proposed in Ref.~\cite{Yoo:2018kvb} 
so that we can decouple the larger-scale environmental effect, 
which is irrelevant to the PBH formation. 
Thus we can eliminate the redundant variance due to the environmental effect, 
and obtain a narrower mass spectrum than that in Ref.~\cite{Yoo:2018kvb}. 
This new procedure also allows us to straightforwardly implement a window function 
and calculate the PBH abundance for an arbitrary power spectrum of the curvature perturbation. 
For a sufficiently narrow power spectrum, the PBH mass spectrum results in 
the case without the window function irrespective of the choice of the window function. 
That is, there is no window function dependence for a sufficiently narrow spectrum in our procedure.

It should be noted that, in Ref.~\cite{Germani:2019zez}, 
the authors attempted to estimate PBH abundance for a broad 
spectrum without a window function. 
In the results in Ref.~\cite{Germani:2019zez}, 
we can find significant enhancement of the mass spectrum in the large-mass region 
compared with our results. 
Although the reason for this discrepancy should be further investigated in the future, 
we make some discussion in Appendix\ref{sec:dis}.

The PBH abundance for 
the scale-invariant flat power spectrum has been calculated in Sec.~\ref{window} 
as an example. 
The result largely depends on the choice of the window function. 
Nevertheless, we found that the $k$-space tophat window function has the minimum required property. 
Specifically, it minimizes the extra reduction of the mass spectrum due to the window function. 
When one estimates PBH abundance without any concrete physical process of the smoothing, 
the choice of the $k$-space tophat window function would be the best in our procedure. 
Finally, we emphasize that our procedure makes it possible to calculate the PBH mass spectrum 
for an arbitrary power spectrum by using a plausible PBH formation criterion 
with the nonlinear relation taken into account.

\section*{Acknowledgements}
We thank Jaume Garriga for his contribution to a part of this work
provided during the discussion in the previous work\cite{Yoo:2018kvb}.  We
also thank Shuichiro Yokoyama 
and Cristiano Germani for helpful comments and stimulating discussion.  
This work was
supported by JSPS KAKENHI Grant Numbers JP19H01895 (C.Y., T.H. and
S.H.), JP19K03876 (T.H.)  and JP17H01131 (K.K.), and MEXT KAKENHI Grant
Nos.~JP19H05114 (K.K.) and JP20H04750 (K.K.).

\appendix

\section{Random Gaussian distribution of $\zeta$}
\label{Gauss}

Due to the random Gaussian assumption, the probability distribution of any set of 
linear combination of the variable $\zeta(x_i)$ is given by a multi-dimensional Gaussian 
probability distribution\cite{1970Afz.....6..581D,1986ApJ...304...15B}: 
\begin{equation}
P(V_I)\dd^n V_I=\left(2\pi\right)^{-n/2}\left|\det \mathcal M\right|^{-1/2}
\exp \left[-\frac{1}{2}V_I \left(\mathcal M^{-1}\right)^{IJ}V_J\right]\dd^n V, 
\end{equation}
where the components of the matrix $\mathcal M$ are given by 
the correlation $<V_I V_J>$ defined by 
\begin{equation}
<V_I V_J>:=\int \frac{\dd \bm k}{(2\pi)^3}\frac{\dd \bm k'}{(2\pi)^3}<\tilde V_I^*(\bm k)\tilde V_J(\bm k')> 
\end{equation}
with $\tilde V_I(\bm k)=\int \dd^3 xV_I(\bm x)\ee^{i\bm k\bm x}$. 

The non-zero correlations between two of $\nu =-\zeta/\sigma_0$, $\xi=\triangle \zeta/\sigma_2$ and 
$\omega=-\triangle \triangle \zeta/\sigma_4$ are 
given by 
\begin{eqnarray}
<\nu\nu>&=&<\xi\xi>=<\omega\omega>=1, \\
<\nu\xi>&=&\gamma_1:=\sigma_1^2/(\sigma_0\sigma_2),\\
<\nu\omega>&=&\gamma_2:=\sigma_2^2/(\sigma_0\sigma_4),\\
<\xi\omega>&=&\gamma_3:=\sigma_3^2/(\sigma_2\sigma_4). 
\end{eqnarray}
Then, the probability distribution function for these variables is given as follows:
\begin{eqnarray}
P(\nu,\xi,\omega)&=&(2\pi)^{-3/2}\left|D \right|^{-1/2}\exp\Biggl[
-\frac{1}{2D}\Bigl\{
(1-\gamma_3^2)\nu^2+(1-\gamma_2^2)\xi^2+(1-\gamma_1^2)\omega^2\cr
&&-2(\gamma_1-\gamma_2\gamma_3)\nu\xi
-2(\gamma_2-\gamma_3\gamma_1)\omega\nu
-2(\gamma_3-\gamma_1\gamma_2)\xi\omega
\Bigr\}\Biggr], 
\label{eq:P}
\end{eqnarray}
where 
\begin{equation}
D=\det \mathcal M=1-\gamma_1^2-\gamma_2^2-\gamma_3^2+2\gamma_1\gamma_2\gamma_3
\end{equation}
with 
\begin{equation}
\mathcal M=
\left(
\begin{array}{lll}
1&\gamma_1&\gamma_2\\
\gamma_1&1&\gamma_3\\
\gamma_2&\gamma_3&1
\end{array}
\right). 
\end{equation}
We re-express the probability $P$ as a probability distribution function $\tilde{P}$ of 
$\zeta_0$, $\mu_2$ and $k_\bullet$, that is 
\begin{equation}
{\tilde{P}}(\zeta_0,\mu_2,k_\bullet)\dd\zeta_0 \dd\mu_2 \dd k_\bullet=
P(\nu,\xi,\omega)\dd \nu \dd \xi \dd \omega
=\frac{2\mu_2k_\bullet}{\sigma_0\sigma_2\sigma_4}P\left(\frac{\zeta_0}{\sigma_0},\frac{\mu_2 }{\sigma_2},\frac{\mu_2 k_\bullet^2}{\sigma_4}\right)\dd\zeta_0 \dd\mu_2 \dd k_\bullet.
\end{equation}
Then, the conditional probability $p(\zeta_0)$ with fixed $\mu_2$ and $k_\bullet$ is given by
\begin{equation}
p(\zeta_0)=\left(\frac{1-\gamma_3^2}{2\pi D \sigma_0^2 }\right)^{1/2}\exp\left[-\frac{1-\gamma_3^2}{2D\sigma_0^2}
\left(\zeta_0-\bar \zeta_0\right)^2\right]
=\left(\frac{1-\gamma_3^2}{2\pi D \sigma_0^2 }\right)^{1/2}\exp\left[-
\frac{1-\gamma_3^2}{2D\sigma_0^2}
\zeta_\infty^2\right], 
\end{equation}
where
\begin{eqnarray}
\bar\zeta_0&=&-\mu_2\frac{\left(\sigma_1^2\sigma_4^2-\sigma_2^2\sigma_3^2\right)+\left(\sigma_2^4-\sigma_1^2\sigma_3^2\right)k_\bullet^2}{\sigma_2^2\sigma_4^2-\sigma_3^4}. 
\label{eq:tilsig}
\end{eqnarray}

\section{Estimation and window function dependence \\in the Press-Schechter formalism}
\label{sec:PS}

For a comparison, we review a conventional estimate 
of the fraction of PBHs based on the PS formalism. 
In the conventional formalism, the scale dependence is introduced by a window function $W(k/k_M)$, 
where 
\begin{equation}
k_M =k_{\rm eq}(M_{\rm eq}/M)^{1/2}. \label{eq:kmM}
\end{equation} 
Then, each gradient moment is replaced by the following expression: 
\begin{equation}
\hat\sigma_n(k_M)^2
=\int\frac{\dd k}{k} k^{2n} \mathcal  P(k) W(k/k_M)^2.  
\label{eq:moments_window}
\end{equation}
The conventional estimate starts 
from the following Gaussian distribution assumption for the density perturbation $\bar \delta$: 
\begin{equation}
P_\delta(\bar \delta)\dd \bar\delta=
\frac{1}{\sqrt{2\pi }\sigma_\delta}\exp\left(-\frac{1}{2}\frac{\bar \delta^2}{\sigma_\delta^2}\right)\dd \bar \delta, 
\label{eq:Gaussian_delta}
\end{equation}
where $\sigma_\delta$ is given by the coarse-grained density contrast 
\begin{equation}
\sigma_\delta(k_M)=\frac{4}{9}\frac{\hat \sigma_2(k_M)}{k_M^2}. 
\label{eq:sigdel}
\end{equation}
Here, for simplicity, we use the same numerical value of $\delta_{\rm th}$ as in our approach, 
in other words, we assume that the volume average of the density perturbation 
obeys the Gaussian probability distribution given by Eq.~\eqref{eq:Gaussian_delta} with 
the coarse-grained density contrast \eqref{eq:sigdel} in the PS formalism. 
This Gaussian distribution and the dispersion are motivated by the linear relation between $\zeta$ and $\delta$.
The fraction $\beta_0^{\rm PS}$ is then evaluated as follows(see e.g. \cite{Carr:2016drx}):
\begin{equation}
\beta_{0}^{\rm PS}(M)=2\alpha \int^\infty_{\delta_{\rm th}}\dd\bar \delta P_\delta(\bar \delta) 
=\alpha
{\rm erfc}\left(\frac{\delta_{\rm th}}{\sqrt{2}\ \sigma_\delta(k_M)}\right)
=\alpha
{\rm erfc}\left(\frac{9}{4}\frac{\delta_{\rm th}k_M^2}{\sqrt{2}\hat \sigma_2(k_M)}\right).
\label{eq:conv0}
\end{equation} 

\begin{figure}[htbp]
\begin{center}
\includegraphics[scale=0.65]{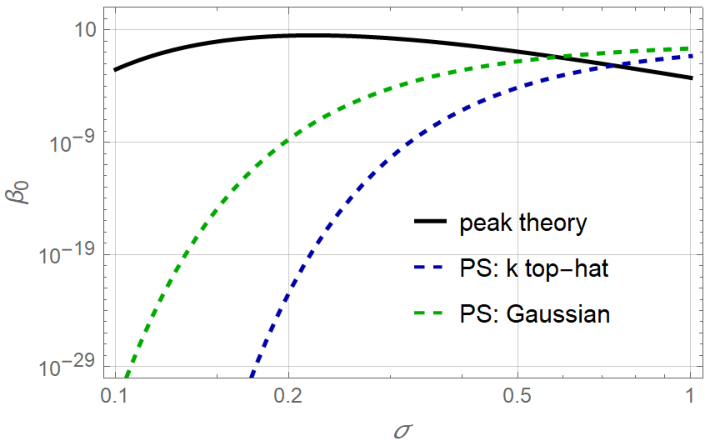}
\includegraphics[scale=0.65]{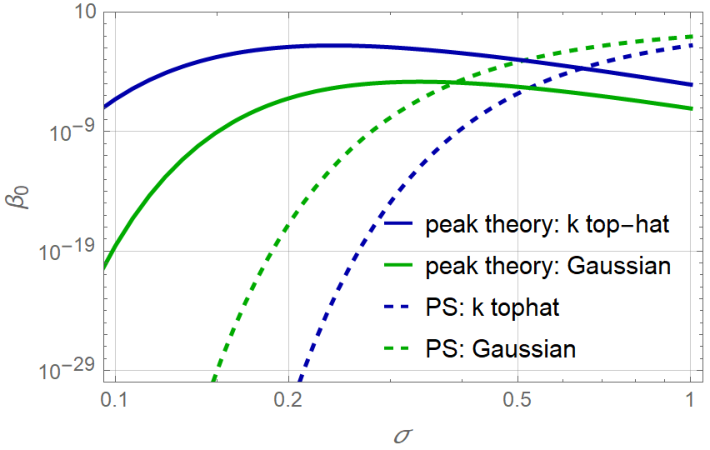}
\caption{
\baselineskip5mm
The window function dependences of the PS formalism and our procedure are 
shown for the monochromatic spectrum~(left panel) and the flat spectrum~(right panel). 
The solid lines and dashed lines show the results for our procedure based on the peak theory and 
for the PS formalism, respectively. 
In the left panel, the values for the PS formalism are given by Eq.~\eqref{eq:beta0PSmono} 
with $k_0=k_M$, and the value in our procedure is depicted as a single 
solid line because the fraction of PBH does not depend on the window function in our procedure. 
}
\label{fig:winddep_PS}
\end{center}
\end{figure}

Let us check the window function dependence in the PS formalism for the cases 
of the monochromatic spectrum $\mathcal P(k)=\sigma^2 k_0\delta(k-k_0)$ and 
the flat spectrum $\mathcal P(k)=\sigma^2$. 
For the monochromatic spectrum, the faction $\beta_{\rm 0,mono}^{\rm PS}$ is given by 
\begin{equation}
\beta_{\rm 0,mono}^{\rm PS}={\rm efrc}\left(\frac{9}{4}\frac{\delta_{\rm th}k_M^2}{\sqrt{2}\sigma k_0^2W(k_0/k_M)}\right). 
\label{eq:beta0PSmono}
\end{equation}
From this expression, the existence and significance of the window function dependence are clear 
(see the left panel in Fig.~\ref{fig:winddep_PS}). 
On the other hand, as is stated in the first paragraph in Sec.~\ref{window}, 
for a narrow power spectrum, there is essentially no window function dependence in our procedure. 
In particular, for the monochromatic spectrum case, the fraction reduces to the result 
given in Ref.~\cite{Yoo:2018kvb} for an arbitrary window function satisfying the properties 
listed in the same paragraph. 

For the flat spectrum, the value of the second gradient moment is given by
$\sigma k_M^2/\sqrt{2}$ and $\sigma k_M^2/2$ for the Gaussian and 
$k$-space tophat window functions, respectively. 
Therefore, in the PS formalism, 
the abundance is larger for the Gaussian window function differently from our case 
shown in Fig.~\ref{fig:winddep}. 
In the right panel of Fig.~\ref{fig:winddep_PS}, 
the window function dependences in the PS formalism and our procedure 
are shown. 
In both cases, the window function dependence is significant at a similar extent. 

It should be noted that in the PS formalism, the reason for the 
larger abundance for the Gaussian window is the 
contribution from the high-$k$ modes through the tail of the Gaussian function. 
Therefore, it is clear that the longer the tail of the window function is, 
the larger abundance becomes. 
Of course, we cannot take the long-tail limit because 
the window function becomes irrelevant in this limit. 
Contrary to the PS formalism, in our procedure, the sharpest cutoff in the $k$-space
gives the largest abundance and the extra reduction due to the window function 
can be minimized in this well-defined limit. 

\section{Discrepancy between our results and those in Ref.~\cite{Germani:2019zez}}
\label{sec:dis}

As is noted in Sec.~\ref{discussion},  
there can be seen a qualitative difference between our PBH mass spectrum 
and that in Ref.~\cite{Germani:2019zez} 
in the large-mass region. 
First, we briefly review the basic idea used in Ref.~\cite{Germani:2019zez}(see also Ref.~\cite{Young:2019yug}). 

In order to clearly distinguish the equations which are valid only for spherically symmetric cases 
from generally valid equations, we use the notation $\circeq$ 
for the equality with spherical symmetry. 
Let us start with the relation between the non-linear volume-averaged density perturbation $\bar \delta$, 
the compaction function $\mathcal C$ and the curvature perturbation\cite{Harada:2015yda}:
\begin{equation}
\bar \delta\circeq2\mathcal C\circeq\frac{2\delta M}{R}\circeq-\frac{2}{3}r_R\zeta_R'[2+r_R \zeta_R'], 
\end{equation}
where $r_R$ is a certain radius and the subscript $_R$ denotes the value at $r=r_R$%
\footnote{
\baselineskip5mm
$r_R$ corresponds to $R$ in Ref.~\cite{Germani:2019zez}.}.  
This equation is valid for super-horizon spherically symmetric perturbations. 
We may define $\delta_{\rm l}$ which is linearly related to $\zeta$ as follows: 
\begin{equation}
\delta_{\rm l}\circeq -\frac{4}{3}r_R\zeta_R'. 
\end{equation}
Then we obtain 
\begin{equation}
\bar \delta\circeq\delta_{\rm l}-\frac{3}{8}\delta_{\rm l}^2. 
\label{eq:deltadelta}
\end{equation}

The linear density perturbation $\delta_{\rm l}$ should be compared with $\delta_R$ defined in 
Eq.~(9) in Ref.~\cite{Germani:2019zez} as follows:
\begin{equation}
\delta_R(r_R)=\frac{3}{4\pi r_R^3}\int d^3x \frac{\delta \rho}{\rho}\theta(r_R-|\vec x-\vec x_0|), 
\label{eq:deltaR}
\end{equation}
where $\theta$ is the Heviside step function, which effectively acts as the real-space tophat window function. 
We note that this expression is defined not only for spherically symmetric perturbations but also for general ones. 
Using the following linear relation:
\begin{equation}
\frac{\delta\rho}{\rho}=-\frac{4}{9}\frac{1}{a^2H^2}\triangle \zeta
\circeq-\frac{4}{9}\frac{1}{a^2H^2}\frac{1}{r^2}\del_r(r^2\del_r \zeta) 
\end{equation}
in a spherically symmetric case, we can find~\cite{Young:2019yug}
\begin{equation}
\delta_R(r_R)\circeq\frac{4}{3r_R}\int^{r_R}_0 dr r^2(\zeta''+\frac{2}{r}\zeta')
\circeq
 -\frac{4}{3}r_R \zeta_R'
\circeq
\delta_{\rm l}
\label{eq:deltas}
\end{equation}
at the horizon entry time defined by $r_R=1/(aH)$. 

In Ref.~\cite{Germani:2019zez}, the relation with 
spherical symmetry $2\mathcal C=\delta_{\rm l}-\frac{3}{8}\delta_{\rm l}^2$
is extended to the general relation $2\mathcal C=\delta_R-\frac{3}{8}\delta_R^2$, 
and the PBH formation criterion for $\mathcal C$ is expressed in terms of $\delta_R$, 
and used to estimate PBH abundance. 
$\delta_R$ is equivalent to the linear density perturbation with the real-space tophat window function. 
However, it should be noted that the real-space tophat window function is 
naturally introduced so that the relation Eq.~\eqref{eq:deltas} can be satisfied, 
and not introduced by hand as a window function.

Let us consider the estimation of PBH abundance 
in the large-mass region where the discrepancy exists. 
For simplicity, let us focus on a single-scale power spectrum with the typical scale $1/k_0$. 
First, we note that, in Ref.~\cite{Germani:2019zez}, 
the value of $r_R$ is chosen such that $\dd \delta_R/\dd r_R=0$ and 
$\mathcal C(\vec x, r_R)$ takes a maximal value at $\vec x=\vec x_0$, 
where  $\mathcal C$ is regarded as a function of $\vec x$ and $r_R$. 
The scale of the relevant region to PBH formation is given by $r_R$, 
which can be significantly different from $1/k_0$%
\footnote{
\baselineskip5mm
In our procedure, the relevant scale is $r_{\rm m}\sim 1/k_0$.  
}.
However, the relevance of the present 
criterion given in terms of the compaction function $\mathcal C$ 
is not clear for the outer maxima.
In Fig.~2 of Ref.~\cite{Atal:2019erb}, 
the result of a numerical simulation for 
a spherically symmetric and oscillatory initial profile is shown. 
The initial profile in the simulation 
corresponds to the most probable profile for the delta-functional 
power spectrum peaked at $1/k_0$. 
The most probable profile is given by a peak at the center surrounded by repeated 
concentric overdense and underdense regions.
More precisely, the most probable profile of the curvature perturbation is given by a sinc 
function, where the compaction function is oscillatory with respect to 
the distance from the center.
We can find that the PBH formation criterion is satisfied 
for the multiple radii corresponding to the maxima in Fig.~2 of Ref.~\cite{Atal:2019erb}. 
The resultant PBH, however, has the mass corresponding to the 
typical scale $1/k_0$, whereas
no PBH of larger mass scales is formed. 
This result suggests that the present criterion is relevant only for the innermost 
maximum of the compaction function but not for the outer maxima.
If the present criterion is simply applied to not only the innermost but also 
outer maxima, the abundance of primordial black holes of large-mass scales 
could be overestimated.


\begin{thebibliography}{10}
\baselineskip5.5mm


\bibitem{Yoo:2018kvb}
C.-M. Yoo, T.~Harada, J.~Garriga, and K.~Kohri,
\newblock PTEP {\bf 2018}, 123E01 (2018), arXiv:1805.03946, {\em {Primordial
  black hole abundance from random Gaussian curvature perturbations and a local
  density threshold}}.

\bibitem{1967SvA....10..602Z}
Y.~B. {Zel'dovich} and I.~D. {Novikov},
\newblock Soviet Ast. {\bf 10}, 602 (1967), {\em {The Hypothesis of Cores
  Retarded during Expansion and the Hot Cosmological Model}}.

\bibitem{Hawking:1971ei}
S.~Hawking,
\newblock Mon. Not. Roy. Astron. Soc. {\bf 152}, 75 (1971), {\em
  {Gravitationally collapsed objects of very low mass}}.

\bibitem{Carr:2020gox}
B.~Carr, K.~Kohri, Y.~Sendouda, and J.~Yokoyama,
\newblock (2020), arXiv:2002.12778, {\em {Constraints on Primordial Black
  Holes}}.

\bibitem{Carr:2020xqk}
B.~Carr and F.~Kuhnel,
\newblock (2020), arXiv:2006.02838, {\em {Primordial Black Holes as Dark
  Matter: Recent Developments}}.

\bibitem{Abbott:2016blz}
Virgo, LIGO Scientific, B.~P. Abbott {\em et~al.},
\newblock Phys. Rev. Lett. {\bf 116}, 061102 (2016), arXiv:1602.03837, {\em
  {Observation of Gravitational Waves from a Binary Black Hole Merger}}.

\bibitem{Sasaki:2016jop}
M.~Sasaki, T.~Suyama, T.~Tanaka, and S.~Yokoyama,
\newblock Phys. Rev. Lett. {\bf 117}, 061101 (2016), arXiv:1603.08338, {\em
  {Primordial Black Hole Scenario for the Gravitational-Wave Event GW150914}}.

\bibitem{DeLuca:2019qsy}
V.~De~Luca {\em et~al.},
\newblock JCAP {\bf 07}, 048 (2019), arXiv:1904.00970, {\em {The Ineludible
  non-Gaussianity of the Primordial Black Hole Abundance}}.

\bibitem{Ando:2018qdb}
K.~Ando, K.~Inomata, and M.~Kawasaki,
\newblock Phys. Rev. D {\bf 97}, 103528 (2018), arXiv:1802.06393, {\em
  {Primordial black holes and uncertainties in the choice of the window
  function}}.

\bibitem{Young:2019osy}
S.~Young,
\newblock Int. J. Mod. Phys. D {\bf 29}, 2030002 (2019), arXiv:1905.01230, {\em
  {The primordial black hole formation criterion re-examined: Parametrisation,
  timing and the choice of window function}}.

\bibitem{Tokeshi:2020tjq}
K.~Tokeshi, K.~Inomata, and J.~Yokoyama,
\newblock (2020), arXiv:2005.07153, {\em {Window function dependence of the
  novel mass function of primordial black holes}}.

\bibitem{Carr:1975qj}
B.~J. Carr,
\newblock Astrophys. J. {\bf 201}, 1 (1975), {\em {The Primordial black hole
  mass spectrum}}.

\bibitem{1978SvA....22..129N}
D.~K. {Nadezhin}, I.~D. {Novikov}, and A.~G. {Polnarev},
\newblock Soviet Ast. {\bf 22}, 129 (1978), {\em {The hydrodynamics of
  primordial black hole formation}}.

\bibitem{1980SvA....24..147N}
I.~D. {Novikov} and A.~G. {Polnarev},
\newblock Soviet Ast. {\bf 24}, 147 (1980), {\em {The Hydrodynamics of
  Primordial Black Hole Formation - Dependence on the Equation of State}}.

\bibitem{Shibata:1999zs}
M.~Shibata and M.~Sasaki,
\newblock Phys. Rev. {\bf D60}, 084002 (1999), arXiv:gr-qc/9905064, {\em {Black
  hole formation in the Friedmann universe: Formulation and computation in
  numerical relativity}}.

\bibitem{Niemeyer:1999ak}
J.~C. Niemeyer and K.~Jedamzik,
\newblock Phys. Rev. {\bf D59}, 124013 (1999), arXiv:astro-ph/9901292, {\em
  {Dynamics of primordial black hole formation}}.

\bibitem{Musco:2004ak}
I.~Musco, J.~C. Miller, and L.~Rezzolla,
\newblock Class. Quant. Grav. {\bf 22}, 1405 (2005), arXiv:gr-qc/0412063, {\em
  {Computations of primordial black hole formation}}.

\bibitem{Polnarev:2006aa}
A.~G. Polnarev and I.~Musco,
\newblock Class. Quant. Grav. {\bf 24}, 1405 (2007), arXiv:gr-qc/0605122, {\em
  {Curvature profiles as initial conditions for primordial black hole
  formation}}.

\bibitem{Musco:2008hv}
I.~Musco, J.~C. Miller, and A.~G. Polnarev,
\newblock Class. Quant. Grav. {\bf 26}, 235001 (2009), arXiv:0811.1452, {\em
  {Primordial black hole formation in the radiative era: Investigation of the
  critical nature of the collapse}}.

\bibitem{Polnarev:2012bi}
A.~G. Polnarev, T.~Nakama, and J.~Yokoyama,
\newblock JCAP {\bf 1209}, 027 (2012), arXiv:1204.6601, {\em {Self-consistent
  initial conditions for primordial black hole formation}}.

\bibitem{Nakama:2013ica}
T.~Nakama, T.~Harada, A.~G. Polnarev, and J.~Yokoyama,
\newblock JCAP {\bf 1401}, 037 (2014), arXiv:1310.3007, {\em {Identifying the
  most crucial parameters of the initial curvature profile for primordial black
  hole formation}}.

\bibitem{Harada:2013epa}
T.~Harada, C.-M. Yoo, and K.~Kohri,
\newblock Phys. Rev. {\bf D88}, 084051 (2013), arXiv:1309.4201, {\em {Threshold
  of primordial black hole formation}},
\newblock [Erratum: Phys. Rev.D89,no.2,029903(2014)].

\bibitem{Harada:2015yda}
T.~Harada, C.-M. Yoo, T.~Nakama, and Y.~Koga,
\newblock Phys. Rev. {\bf D91}, 084057 (2015), arXiv:1503.03934, {\em
  {Cosmological long-wavelength solutions and primordial black hole
  formation}}.

\bibitem{Escriva:2019phb}
A.~Escriv\`a, C.~Germani, and R.~K. Sheth,
\newblock Phys. Rev. {\bf D101}, 044022 (2020), arXiv:1907.13311, {\em
  {Universal threshold for primordial black hole formation}}.

\bibitem{Escriva:2020tak}
A.~Escriv\`a, C.~Germani, and R.~K. Sheth,
\newblock (2020), arXiv:2007.05564, {\em {Analytical thresholds for black hole
  formation in general cosmological backgrounds}}.

\bibitem{Germani:2019zez}
C.~Germani and R.~K. Sheth,
\newblock Phys. Rev. D {\bf 101}, 063520 (2020), arXiv:1912.07072, {\em
  {Nonlinear statistics of primordial black holes from Gaussian curvature
  perturbations}}.

\bibitem{Suyama:2019npc}
T.~Suyama and S.~Yokoyama,
\newblock PTEP {\bf 2020}, 023E03 (2020), arXiv:1912.04687, {\em {A novel
  formulation of the PBH mass function}}.

\bibitem{1986ApJ...304...15B}
J.~M. {Bardeen}, J.~R. {Bond}, N.~{Kaiser}, and A.~S. {Szalay},
\newblock \apj {\bf 304}, 15 (1986), {\em {The statistics of peaks of Gaussian
  random fields}}.

\bibitem{Musco:2012au}
I.~Musco and J.~C. Miller,
\newblock Class. Quant. Grav. {\bf 30}, 145009 (2013), arXiv:1201.2379, {\em
  {Primordial black hole formation in the early universe: critical behaviour
  and self-similarity}}.

\bibitem{Yoo:2019pma}
C.-M. Yoo, J.-O. Gong, and S.~Yokoyama,
\newblock JCAP {\bf 09}, 033 (2019), arXiv:1906.06790, {\em {Abundance of
  primordial black holes with local non-Gaussianity in peak theory}}.

\bibitem{Choptuik:1992jv}
M.~W. Choptuik,
\newblock Phys. Rev. Lett. {\bf 70}, 9 (1993), {\em {Universality and scaling
  in gravitational collapse of a massless scalar field}}.

\bibitem{Koike:1995jm}
T.~Koike, T.~Hara, and S.~Adachi,
\newblock Phys. Rev. Lett. {\bf 74}, 5170 (1995), arXiv:gr-qc/9503007, {\em
  {Critical behavior in gravitational collapse of radiation fluid: A
  Renormalization group (linear perturbation) analysis}}.

\bibitem{Niemeyer:1997mt}
J.~C. Niemeyer and K.~Jedamzik,
\newblock Phys. Rev. Lett. {\bf 80}, 5481 (1998), arXiv:astro-ph/9709072, {\em
  {Near-critical gravitational collapse and the initial mass function of
  primordial black holes}}.

\bibitem{Yokoyama:1998xd}
J.~Yokoyama,
\newblock Phys. Rev. {\bf D58}, 107502 (1998), arXiv:gr-qc/9804041, {\em
  {Cosmological constraints on primordial black holes produced in the near
  critical gravitational collapse}}.

\bibitem{Green:1999xm}
A.~M. Green and A.~R. Liddle,
\newblock Phys. Rev. {\bf D60}, 063509 (1999), arXiv:astro-ph/9901268, {\em
  {Critical collapse and the primordial black hole initial mass function}}.

\bibitem{Kuhnel:2015vtw}
F.~Kuhnel, C.~Rampf, and M.~Sandstad,
\newblock Eur. Phys. J. {\bf C76}, 93 (2016), arXiv:1512.00488, {\em {Effects
  of Critical Collapse on Primordial Black-Hole Mass Spectra}}.

\bibitem{Germani:2018jgr}
C.~Germani and I.~Musco,
\newblock Phys. Rev. Lett. {\bf 122}, 141302 (2019), arXiv:1805.04087, {\em
  {Abundance of Primordial Black Holes Depends on the Shape of the Inflationary
  Power Spectrum}}.

\bibitem{1970Afz.....6..581D}
A.~G. {Doroshkevich},
\newblock Astrofizika {\bf 6}, 581 (1970), {\em {The space structure of
  perturbations and the origin of rotation of galaxies in the theory of
  fluctuation.}}

\bibitem{Carr:2016drx}
B.~Carr, F.~Kuhnel, and M.~Sandstad,
\newblock Phys. Rev. {\bf D94}, 083504 (2016), arXiv:1607.06077, {\em
  {Primordial Black Holes as Dark Matter}}.

\bibitem{Young:2019yug}
S.~Young, I.~Musco, and C.~T. Byrnes,
\newblock (2019), arXiv:1904.00984, {\em {Primordial black hole formation and
  abundance: contribution from the non-linear relation between the density and
  curvature perturbation}}.

\bibitem{Atal:2019erb}
V.~Atal, J.~Cid, A.~Escriv\`a, and J.~Garriga,
\newblock JCAP {\bf 05}, 022 (2020), arXiv:1908.11357, {\em {PBH in single
  field inflation: the effect of shape dispersion and non-Gaussianities}}.

\end{thebibliography}

\end{document}